\documentclass[twocolumn]{article}
\def\ldw{\vskip 10pt}
\def\itl{\'{\i}}
\def\etal{{\it et al.~}}

\def\half  {\textstyle {1 \over 2} \displaystyle}
\def\2ov3  {\textstyle {2 \over 3} \displaystyle}
\def\3ov2  {\textstyle {3 \over 2} \displaystyle}
\def\5ov2  {\textstyle {5 \over 2} \displaystyle}
\def\11ov12  {\textstyle {11 \over 12} \displaystyle}

\def\beq{\begin{equation}}
\def\eeq{\end{equation}}

\def\gam{{$\gamma$}}

\def\msun{{\rm ~M}_\odot}
\def\AMp{{ {\rm A}^{\rm eff}_{_{\cM +}} }}
\def\AM{{ \hat{\rm A} }}

\def\cB{{\cal B}}
\def\cP{{\cal P}}
\def\cM{{\cal M}}
\def\bh{{\rm bh}}
\def\BH{{\rm BH}}
\def\cH{{\cal H}}
\def\cK{{\cal K}}
\def\cL{{\cal L}}
\def\cO{{\cal O}}
\def\cS{{\cal S}}
\def\cT{{\cal T}}
\def\Kep{{\rm Kep}}

\def\Valf{{\it v}^{\rm Alf}}
\def\Alf{{\rm Alfv\'en}~}
\def\Oor{{\rm Oort}}
\def\SS{{\rm SS}}

\def\Prad{p_{\rm rad}}
\def\Pgas{p_{\rm gas}}
\def\Pb{p_{\rm\bf b}}
\def\PB{p_{\rm\bf B}}

\def\bUps{ {\mbox{\boldmath $\Upsilon$}} }
\def\bPsi{ {\mbox{\boldmath $\Psi$}} }
\def\bxi { {\mbox{\boldmath $\xi$}} }
\def\unvc{ {\mbox{\boldmath $1$}} }

\def\be{\begin{equation}}
\def\ee{\end{equation}}

\def\<{{\langle}}
\def\>{{\rangle}}

\def\sA{{\mathcal{A}}}

\def\sD{{\mathcal{D}}}
\def\sE{{\mathcal{E}}}

\def \ltaprx {\lower .1ex\hbox{\rlap{\raise .6ex\hbox{\hskip .3ex
	{\ifmmode{\scriptscriptstyle <}\else 
		{$\scriptscriptstyle <$}\fi}}}
	\kern -.4ex{\ifmmode{\scriptscriptstyle \sim}\else 
		{$\scriptscriptstyle\sim$}\fi}}}
\def\gtaprx {\lower .1ex\hbox{\rlap{\raise .6ex\hbox{\hskip .3ex
	{\ifmmode{\scriptscriptstyle >}\else 
		{$\scriptscriptstyle >$}\fi}}}
	\kern -.4ex{\ifmmode{\scriptscriptstyle \sim}\else 
		{$\scriptscriptstyle\sim$}\fi}}}

\def\sec{{\rm ~sec}}

\begin{document}

\title{Hydromagnetic Energy Conversion and Prompt Collimation \\
	in Mildly Advective, Kerr Black 
Holes\footnote{ 
submitted to {\it M.N.R.A.S.},  astro-ph/9912324}
}

\author {
 	Rafael A. Araya-G\'ochez 	\\
  	{\small Laboratorio de Investigaciones Astrof{\itl}sicas,
	Escuela de F{\itl}sica} 	\\  
  	{\small Universidad de Costa Rica, San Jos\'e, Costa Rica.}
  	}


\date{}
\maketitle

\begin{abstract} 
	Recent evidence of the phenomenal energetics involved in \gam-ray 
 burst events strongly suggests that the progenitor must efficiently convert
 gravitational binding energy into a moderately collimated outflow,
 possibly in the form of a Poynting jet.
 We show that an MHD-instability driven dynamo (IDD) operating
 in a hot accretion disk is capable of generating energetically adequate
 magnetic flux deposition rates above and below a 
 mildly advective accretion disk structure.  		
 The dynamo is driven by the magnetorotational instability (MRI) of a 
 toroidal field in a shear flow and is limited by the buoyancy of 
 `horizontal' flux and by reconnection in the turbulent medium.  
 In the comoving frame of a semi-thick, slim disk setting, the predominant
 field components reside in surfaces perpendicular to the local meridian
 and the flux is deposited in funnels that are relatively free of baryons. 
 The efficiency of magnetic energy deposition is estimated to be comparable
 to the neutrino losses but the strong effective shear induced by the metric 
 on the MRI favors pumping magnetic field energy at low wavenumbers, 
 i.e., field generation at large coherence lengthscales.  This, in turn,
 suggests that an MHD collimation mechanism may deem this process a more
 viable alternative to neutrino-burst--driven models of \gam-ray bursts.
\end{abstract}    


\section{Introduction}
 \label{sec:Intro}

	The combined redshift and fluence measurements
 of at least five \gam-ray burst sources: GRB970508 @ $z=.835$ 
 (Metzger \etal 1997, Kouveliotou \etal 1997), GRB971214 @ $z=3.4$
 (Kulkarni \etal 1998, Kippen \etal 1997), GRB980718 @ $z=.966$ 
 (Djorgovski \etal 1998, Kippen \etal 1998), GRB990123 @ $z= 1.6$
 (Kelson \etal 1999, Kippen 1999a, Conners \etal 1999), and
 GRB990510 @ $z=1.619$ (Vreeswijk \etal 1999,  Kippen \etal 1999b),
 plus very large photon energy detections in certain bursts
 (e.g. Sommer \etal 1994, Hurley \etal 1994, Dingus 1995)
 and tight size constraints derived from the rapid risetimes of burst 
 triggers (Walker, Schaefer \& Fenimore 1998) strongly suggest that
 the release of energy is highly focused by the central
 engine that propels a \gam-ray burst.

	In spite of the very large \gam-ray energy
 requirements, $E_\gamma \simeq \zeta \times 1.0_{+54}$ erg 
 ($\zeta \equiv \delta \Omega_\gamma / 4\pi$),
 the efficiency of energy deposition into 
 electromagnetic channels is likely to be very poor, 
 $E_\gamma/E_{\rm tot} \equiv \varepsilon \ltaprx [.01, .001]$, 
 ~if the burst is driven by a neutrino burst in analogy to the processes
 thought to give birth to supernov\ae\, (MacFadyen \& Woosley 1998, 1999)
 or if the burst involves major energy losses to gravitational radiation
 such as might be the case in compact object merger scenarios 
 (Rasio and Shapiro 1992, 1994, Davies \etal 1994, 
 Janka \& Ruffert 1996, 1998; Ruffert \& Janka 1998).

	Thus, although the monumental observational progress
 of late at first prompted some workers to suggest that 
 ``the physical mechanisms behind \gam-ray bursts (were) within 
 reach" (Metzger \etal 1997), 
 the unexpectedly large increase in the energy and/or collimation
 requirement of any viable \gam-ray burst progenitor has deemed such 
 mechanisms at present largely undetermined.  Indeed these measurements
 pose a serious energy budget problem for arguably all gravitational 
 collapse powered
 theoretical models of \gam-ray bursts if the energy release is not 
 moderately collimated $\zeta \ltaprx~1.0_{-3}$  
 (see, however, M\'esz\'aros, Rees \& Wijers 1999).

	On the other hand, the angular spreading problem (Fenimore, Epstein
 \& Ho 1996, Fenimore, Mandras \& Nayakshin 1996) means that it may be quite
 implausible for complex millisecond $\gamma$-ray variability to be 
 attributed to the interaction of a single fireball with an external medium.
 The somewhat favored scenario is one in which the variability down to
 possibly sub-ms timescales portrays flux fluctuations at the
 {\it emission radius} $R_e$ (Sari \& Piran 1997, Kobashi, Piran 
 \& Sari 1998) or from flaring within the energy source itself. 
 Fenimore, Ramirez-Ruiz \& Wu (1999) assert that the latter must be the 
 case for the brightest burst yet: GRB990123. It is not clear, however, 
 how such radiation could avoid thermalization unless the energy
 release mechanism yields infinitesimally thin shells 
 (M. Baring, Priv. Comm.).   

 	An attractive solution to the energy budget/collimation {\it problem}
 starts with a directional Poynting-flux dominated outflow 
 (Thompson 1994,  M\'esz\'aros \& Rees 1997) under the premise 
 that such a flow may carry very little baryonic contamination 
 if deposited along a centrifugally (or gravitationally) evacuated
 funnel such as the angular momentum axis of a black hole-accretion disk 
 system.  

	The notion that magnetic fields may play an important role in tapping 
 the energy available in accretion disks and in its subsequent re-processing 
 into a high Lorentz factor outflow has been surmised by several authors. 
 Attempts have been made at drawing an analog to the standard phenomenology
 of pulsar electromagnetic emission (e.g, the magnetized differential rotor
 Katz 1997) and at invoking the Blandford-Znajek mechanism (BZm) 
 (Paczy\'nski 1998) and BZm-like processes at the innermost regions of accretion
 disks (Ghosh \& Abramowicz 1997, Livio, Ogilvie \& Pringle 1998). 

 	These estimates of Poynting flux luminosities assume
 {\it a priori} the existence of a nearly equipartition, {\it large-scale}
 ($\cO [r])$, external magnetic field especially with regards to jet 
 formation processes (Blandford \& Payne 1982, Tsinganos \& Sauty 1994,
 Scheuer 1994, Blandford 1994) even if the magnetic collimation
 mechanism does not explicitly involve significant baryon entrainment 
 (Lynden-Bell 1997,  Lynden-Bell \& Boily 1994, Appl \& Camenzind 1993).  
 Yet, such an assumption is hard to justify on theoretical grounds. 
 On the one hand, self-generated magnetic fields in 
 thin Keplerian disks are unlikely
 to attain coherence lengthscales far in excess of the disk's pressure
 scale height, $\cH$, and inverse-cascade estimates in thin disks 
 (Tout and Pringle 1995) indicate that the large scale component is much
 weaker $\cO (\cH/r)$.  On the other hand, advection of
 a large scale fields from the outer disk or stellar mantle is
 unlikely to permit a substantial steady inflow of material 
 (Ghosh \& Abramowicz 1997) and thus preclude the large 
 accretion rate required by the overall energetics.

	In addition, jet-launching by large scale magnetic fields is known
 to possess a large degree of baryon entrainment (see, e.g. Pringle 1993). 
 This `problem' (from the \gam-ray burst standpoint) persists even 
 when the collimation mechanism does not invoke the presence of a baryonic 
 wind (e.g. Lynden-Bell \& Boily 1994,  Lynden-Bell 1997)
 because a large scale field connects 
 baryon-free with baryon-loaded regions of the system.   Thus, just how the
 Poynting jet might form is an outstanding theoretical problem posed to
 remain a subject of controversy and future research for some time.    

	On the other hand, to our knowledge formal motivation for an 
 external field with the desired strength has yet to be investigated
 in this setting (with the possible exception of Thompson (1994) who
 invokes magnetar-like processes, i.e. strong neutrino induced convective
 instabilities at near nuclear densities).
 Simple `winding-up' of radial field to produce a toroidal
 component was invoked by Narayan, Paczy\'nski \& Piran (1992) and 
 by Katz (1997) but this mechanism does not replenish the radial field
 thus failing to operate as a dynamo.  Parker ($\cP$-type) and 
 magnetorotational instabilities ($\cM$-type or MRI's for short) 
 are also often invoked without further elaboration on how this processes
 might operate together under the conditions implied by \gam-ray burst
 models (see, however, Tout \& Pringle 1992, hereafter TP92, for a general, 
 quantitative account of an instability driven dynamo).

	We envisioned the accretion disk setting following the standard
  hyper-accreting black hole model of Popham, Woosley \& Fryer (1999, 
  hereafter PWF): $M_\bh = 3 \msun; ~\alpha_\SS = 0.1,
 ~{\rm and}~ \dot{M} = 0.1 \msun \sec^{-1}$. 
 These authors find that the onset of photodisintegration and of neutrino
 cooling at radii $r \ltaprx \,70 ~[GM/c^2]$ yield a mildly advective
 accretion disk 
structure\footnote{
 Here the scale height, $\cH_\theta = r \Theta_\cH$,
 corresponds to an average on spherical surfaces
 consistent with a not so thin, partially advective disk 
 (see, e.g., Abramowicz, Lanza \& Percival 1997).
} ($\cH_\Theta / r \ltaprx .4$)
 with semi-Keplerian rotation rate.  The pressure at $r \simeq 70$ is dominated
 by the nucleon gas component and inside this radius, it becomes dominated 
 by radiation at $r \ltaprx~ 40$.  The temperature rises above 
 ($1\gamma$) {\it pair creation threshold} at $r \simeq 56$.  
 For a viscosity parameter $\alpha_\SS = .1$,
 a fluid element's total number of windings is very limited, 
 $N_{\rm wind} \approx 10$, but this increases almost tenfold
 by reducing $\alpha_\SS$ by the same factor.   
 The total neutrino luminosity, about 2\% of $\dot{M}c^2$, and 
 annihilation rate efficiency, $\approx 12\%$, for this model are probably
 insufficient to explain moderately bright \gam-ray bursts but the
 authors remark that magnetohydrodynamic processes might be more 
 efficient at producing a high energy (Poynting) jet.  

	We elaborate on this suggestion using an instability-driven 
 accretion disk dynamo ({\bf IDD}) mechanism.   Similar versions of
 this model (Tout \& Pringle 1992, hereafter TP92) involving 
 $\cP$-type and axisymmetric $\cM$-type (i.e. Balbus-Hawley) 
 instabilities have meet with limited 
 success in explaining dwarf nov\ae\, quiescence and eruption phenomenology 
 (see Armitage, Livio \& Pringle 1996 and references therein).  However, 
 invoking Parker's undulate instability to promote the growth (loss) of
 vertical (horizontal) field is questionable in turbulent disks where the
 turbulence is fed by MRI instabilities (Vishniac \& Diamond 1992),
 unless the growth and length scales of $\cP$-modes could compare 
 favorably to the MRI scales (see \S \ref{ssec:Pair_Eff}). 
 It is at present unclear what role, if any, $\cP$ modes may play 
 in the present context where the (presumably) thermal pairs are
 collisionally well coupled to the nucleon fluid.   
 Following this argument, $\cP$-modes are assumed 
 to be suppressed thus precluding the meridional
 field from becoming dynamically significant.
 By contrast, MRI's will develop at the largest lengthscales permitted by
 the local field topology in spite of the turbulent component to the 
 flow/field.  The field lines should be, in fact, only mildly
 stochastic on the largest MRI lengthscale.

	The dynamics of a dynamo process occurring inside an accretion 
 disk are likely to be fairly sensitive to the equilibrium field structure
 and topology.  
 Because  MRI's are surely present in any highly conductive
 astrophysical disk, no well defined global field topology
 is expected (inside the disk). 
 In addition, the field structure could be either diffusive 
 or intermittent, i.e. concentrated in semi-empty flux ropes.
 In a recent paper (Araya-G\'ochez 2000a), we have looked 
 into the case of an intermittent field in a static metric.  
 Here we consider both cases in a more general stationary metric. 

 	The field components correspond to 
 {\it azimuthally averaged} quantities in the comoving frame, i.e. in
 the local standard of rest frame.  These are consistent with radial
 field generation
 from the MRI evolution of a toroidal magnetic field (i.e. non-axisymmetric
 $\cM$-type instability) which is, in turn, generated by the sheared radial
 field; with both components limited by the buoyancy of nearly `horizontal'
 flux and by fast turbulent reconnection of field lines.  This model is only
 weakly sensitive to the details of the turbulent state but it does depend 
 explicitly on the magnetic Mach number of the turbulence.
 
	Although MHD disk simulations with vertical stratification 
 generally fail to support the buoyancy of magnetic fields
 (Brandenburg, Nordlund, Stein \& Torkelsson 1995;
 Stone, Hawley, Gammie \& Balbus 1996), these results prove to be very 
 sensitive to the choice of vertical boundary conditions and do not account
 for the effect of radiation/pairs (see \S \ref{ssec:Pair_Eff}). 

 	Lastly, we note that the field estimates are not explicitly 
 sensitive to the precise value of the viscosity parameter, $\alpha_\SS$, 
 but depend rather on local pressure ratios, on the relativitic 
 generalization of the shear parameter
 (A$^{\rm Rel}_\Oor$ = Oort's A constant), and 
 on the linear timescale for development of the non-axisymmetric MRI
 (which is found to differ from A$^{\rm Rel}_\Oor$
 by a general relativistic correction factor, \S A).
 Angular momentum may be transported by the ordered and/or the turbulent
 components to the field but the relative contributions to the viscosity
 are not clear at present.

	A review of the slim disk setting in the Kerr geometry 
 and associated disk thermodynamics is given in \S \ref{sec:DiskSett}.  
 The instability driven dynamo in a mildly advective disk is laid out 
 in \S \ref{sec:IDD}.  We then critique and review the assumptions,
 find steady state solutions to the dynamo equations and use these
 to estimate efficiency of the hydromagnetic energy deposition from field 
 buoyancy in \S \ref{sec:toying}.

\section{Accretion Disk Setting}
 \label{sec:DiskSett}

\subsection{Slim disk in Kerr geometry}
 \label{ssec:SlimKerr}

 	The Boyer-Lindquist generalization of the Schwarzschild coordinates
 $(t, ~r, ~\theta, ~\varphi)$ affords the most popular form of the metric
 for a rotating Kerr black hole.  In normalized geometrical units 
 ($c = G = M_\bh = 1$),
 the angular momentum of the hole is $a = {\cal J} ~~[1/(GM_\bh/c^2)]$. 
 
	The metric functions of the radial {\bf BLF} coordinate are written
 as relativistic corrections 
(e.g. Novikov and Thorne 1973)\footnote{
 A useful relation connecting $\sA$ and $\sD$
 is:  $~(1-2/r)\sA + 4a^2/r^4 = \sD$.
}:
\beq
\sA \equiv 1 + a^2/r^2 + 2 a^2/r^3,
~~{\rm and}~~
\sD \equiv 1 - 2/r + a^2/r^2,
\eeq so that,
 in the equatorial plane, the squared line-element may be written as: 
\beq
ds^2 = -{\sD\over{\sA}} dt^2 + r^2\sA(d\phi - \omega d t)^2
        + {1\over{\sD}}dr^2,
\label{eq:Metric} \eeq
 with $\omega \equiv {2 a / {\sA r^3}}$
 the rate of frame dragging by the hole.  

	At $\theta = \pi/2$, the horizon lies at 
 $r_+ = 1 + \sqrt{1 - a^2}~(\sD|_{r_\pm}=0)$
 and no observers with time-like worldlines may keep
 $\varphi$ constant inside the static limit at $r_{\rm S} = 2$.

	The Killing vector fields that belong to the azimuthal 
 and time symmetries of the metric are deliberately simple in the BLF: 
 $\bPsi = (0,0,0,1) ~~{\rm and}~~ \bUps = (1,0,0,0)$.  These may be used
 to show that the time and azimuthal components of the four-velocity, 
 $(U_t,U_r,U_\theta,U_\varphi)$, are conserved along geodesics.  
 We denote the latter according to standard notation: 
 $U_t = p_t \equiv -\sE$, and  $U_\varphi = p_\varphi \equiv \ell$.   

	The unit time-like vector is constructed
 from the killing vectors: $\unvc_{t} \propto \bUps + \omega \bPsi$.
 This expression defines ``forwards in time" everywhere in the flow.  
 The rate of frame dragging and the radius of gyration for circular
 orbits are quantities derived from inner products of Killing vector 
 fields (Abramowicz, Chen, Granath \& Lasota 1997): 
 $ \omega = -(\bUps \bPsi) / (\bPsi \bPsi)$ and 
 $ \varpi = (\bPsi \bPsi) / |\bUps + \omega \bPsi|$, respectively.

	A natural frame to treat the local aspects of disk flow is the 
 locally non-rotating ({\bf LNR}) frame which is dragged by the hole at
 a rate $\omega$ with respect to distant observers and where 
 $\unvc_{\tilde{t}} = (1,0,0,0)$.   
 Bardeen, Press and Teukolsky (1972) pointed out the advantages of
 calculating physical processes there instead of in the Boyer-Lindquist
 frame which is fixed with respect of distant stars.
 However, in the spirit of the shearing sheet 
 approximation used in the next section, the frame
 we choose to evaluate the ingredients necessary for dynamo action
 is the local rest frame ({\bf LRF}) of the fluid where, in general, 
 $\unvc_{\hat{t}} \neq (1,0,0,0)$ (in our notation $\tilde{t}$
 denotes the time component measured in the LNR frame while $\hat{t}$
 denotes the same measured in the LRF).

	The flow moves with respect to the LNR frame with azimuthal 
 three speed $v^\varphi = \dot{x}^{\tilde{\varphi}} / \dot{x}^{\tilde{t}} 
 ~(\equiv \beta_\varphi)$.  This is related to the angular velocity,
 $\Omega$, as measured by distant observers by
\beq
 v^\varphi 	= \varpi \times (\Omega - \omega) 
		= r ~ {\sA \over {\sqrt{\sD}}} ~ 
		\left( { {\dot{x}^{\varphi} \over \dot{x}^{t}} } \right)
		~~~  \left( = { \ell \over {r \gamma \sqrt{\sA}} } \right).
\label{eq:GyrRad} \eeq

	From the frame that corotates {\it with} 
 the fluid, the measured radial velocity is defined as $\beta_r$.   
 This choice of frame to measure the radial velocity dictates that 
 $\beta_r \leq 1$ everywhere in the flow (reaching the upper limit 
 at $r_+$) and that $\beta_r \approx c_s$ at the sonic surface
 (Abramowicz \etal 1997). 

 	The orthonormal thetrad of basis LRF vectors, used to transform 
 BLF $\Longleftrightarrow$ LRF, and explicit forms
 for the four velocity as function of $\beta_r$ and $\ell$ in the BLF 
 are given in Gammie and Popham (1998, hereafter GP98). 
 For completeness we exhibit the contravariant components 
 (i.e. the momenta) which are easily derivable from the above definitions
 and from $\unvc_t^\nu p_\nu = -\gamma$: 
\beq
	\left(U_t, ~U_r, ~U_\theta, ~U_\varphi \right)  = 
	\left( -\gamma \sqrt{\sD\over{\sA}} - \ell\omega, 
		~{\gamma_r \beta_r \over{\sqrt{\sD }}}, ~0, ~\ell \right), 
\eeq
 where $\gamma = \gamma_\varphi \gamma_r$,
 with $\gamma_\varphi$ and $\gamma_r$ as measured in the frames 
 indicated above.


	By symmetry and for
 simplicity $U^\theta$ is assumed to vanish at the equatorial plane 
 and periodic azimuthal boundary conditions in the {``\it dragged"} frame
 allow for non-axisymmetric unstable MHD modes to develop while permitting
 axisymmetry of the mean flow.   Although the equations governing the flow 
 are greatly simplified by the symmetry conditions, a potential caveat of 
 this assumption is that low wavenumber non-axisymmetric unstable azimuthal
 modes, such as undulate $\cP$-type modes, may violate these conditions
 if the mean-fields become kinematically important.

	In accordance with the standard (one-dimensional) slim-disk 
 approach (Abramowicz \etal 1989, Narayan \& Yi 1995ab, GP98), 
 the `vertical' average is performed over
 the meridional and azimuthal coordinates
\beq
	\int \!\!  \sqrt{g} \, d\phi d\theta ~ f 
		~\equiv~ r^2 \, 2\pi \, 2\Theta_\cH ~ f |_{\theta = \pi/2},
\label{eq:VertInt} \eeq
 where: $f$ is any dynamical variable, $\Theta_\cH = r^{-1}\cH_\theta$ 
 is the meridional angular pressure scale height of the disk (formally
 defined below) and where 
\[	g \equiv  |{\rm Det}(g_{\mu\nu})| 
		= (r^2 + a^2\cos^2\theta) ~r^2 \sin^2\theta,
\]
 is the determinant of the full metric.  

	With $T^{\mu\nu} = 
 p g^{\mu\nu} + \varrho u^{\mu} u^{\nu} + t_{\rm visc}^{\mu\nu}$,
 the viscous stress-energy tensor, four vertically averaged equations
 for $~\ell, ~\beta_r, ~T, ~\rho$ and for the angular momentum eigenvalue,
 $j$, may be derived from particle number, $(\rho u^\mu)_{;\mu} = 0$, and 
 energy-momentum, $ {T^{\mu\nu}}_{;\nu} = 0$, conservations 
 (GP98, PWF, Abramowicz \etal 1997).
 In \S A, we will use the radial momentum conservation equation
\beq
	\gamma_r^2 \beta_r {d \beta_r\over{d r}} =  
		-{1\over{r^2}}{\sA \gamma_\phi^2\over{\sD}}
		\left(1 - {\Omega\over{\Omega}}_+\right) 
		\left(1 - {\Omega\over{\Omega}}_-\right)
		- {1\over{\varrho}} {d p\over{d r}},			
\label{eq:SlimKerr} \eeq
 where $p$ is the total pressure (including the magnetic contribution,
 e.g. \S \ref{ssec:FluiCoup}),  
 $\varrho = \rho + u + p$ is the (unnormalized) specific enthalpy and 
 $\Omega_\pm = \pm (r^{3/2} \pm a)^{-1}$ are the prograde 
 and retrograde Keplerian rotation frequencies of circular, planar orbits
 as measured in the BLF. 

  	PWF find that cooling from neutrino losses and from 
 photodisintegration of He nuclei becomes important as the 
 temperature and density rise.  The cooling is calculated 
 explicitly (i.e. by solving the global ADAF).
 The disk is found to be optically thin to neutrinos for
 $\dot{M} \ltaprx 1.0 \msun \sec^{-1}$, so that neutrino producing
 processes, pair annhilation and pair capture on nuclei,
 act to cool the medium at moderate accretion rates.  

	Where the mass fraction of nucleons $X_{nuc} < 1$, 
 photodisintegration quickly cools off the disk.  
 At moderate accretion rates, $\dot{M} \gtaprx 0.1 \msun \sec^{-1}$,
 photodisintegration is complete at radii $r \ltaprx \, 70$
\footnote{
 The nucleonalization radius should move farther out with the spin parameter,
 $a$, of the hole as it tends to increment the density at constant temperature
 (see next footnote).  Evidently, higher accretion rates move this radius
 outwards as well.
}.  The nucleon component may then be
 assumed to behave like an ideal gas; decoupled from the radiation field
 except to maintain an equilibrium temperature with the thermal bath 
 (no two-temperature model is expected as in the optically thin case).  

\subsection{Fluid component coupling \\ and effective sound speed}
 \label{ssec:FluiCoup}

	Ignoring electron degeneracy pressure,
 the equation state for the medium has two separate contributions 
 $p = \rho T + a' T^4$, where 
 $a' = \11ov12 \, a_{\rm rad}$ accounts for radiation and pairs.
 The effect of degeneracy in the equation of state is found (by PWF)
 to be minimal for $\dot{M} \leq 0.1 \msun \sec^{-1}$.
 Thus, unless the cooling is significantly enhanced; e.g. if 
 the dynamo constitutes a significant sink of energy from
 the shear flow, one may 
cautiously\footnote{
 PWF find that when the rotation parameter of the hole is large, 
 $a \simeq .95$, this results in substantial enhancement of the density
 at the innermost region of the disk ($r \ltaprx 10$).  An accompanying 
 increase in the efficiency of neutrino cooling then acts as a thermostat 
 precluding the temperature from rising.  Whether this situation makes the
 degeneracy pressure contribution to the equation of state non-negligible
 is unclear though. 
} neglect electron degeneracy in the disk.   	

	With the above approximation, the nucleon-gas+radiation
 medium is expected to posses an effective adiabatic index
 ranging from that of a pure relativistic radiation-$e^\pm$ pair
 gas to that of a non-relativistic, ideal, heavy, nucleon gas 
 $\Gamma \equiv ( \partial_{\, [\ln \rho]} \ln p )_s \in [4/3,~5/3]$.
 The pressure from an isotropically tangled (i.e. turbulent) magnetic
 field is included by noting that such field couples well to the (heavy)
 nucleon-gas component through flux freezing, even though the turbulent
 field component by itself would be expected to behave like a relativistic
 gas (e.g. Narayan \& Yi 1995b, Quataert \& Narayan 1999).
 The coherent and turbulent component contributions to the pressure are
 $\PB^{\rm cohe} \equiv  \PB = B^2/8\pi$, and
 $\PB^{\rm turb} \equiv  \Pb = b^2/24\pi$ respectively, 
 with {\bf B}$_{\rm local}$ = {\bf B + b}.

	Defining $\beta \equiv \Prad / (\Pgas + \Pb)$, 
 and assuming a relativistic thermal $e^\pm$ pair distribution function, 
 the relativistic sound speed is simply 
 (Chandrasekhar 1939, Mihalas \& Mihalas 1984)
\beq
	c_s^2 	= \Gamma \, {p \over{\varrho}}, ~~~~{\rm with}~~ 
	\Gamma 	= \left( \frac	{\5ov2 + 20\beta + 16\beta^2}	
				{(\3ov2 + 12\beta)(1 + \beta)} \right).
\eeq
 Conversely, were there a significant fraction of pairs populating 
 suprathermal energies in a less dense environment, the former would
 couple to the {\it mean-field} component through Parker's (1965) 
 suprathermal MHD mode thus yielding two natural sound speeds 
 for the medium (\S \ref{ssec:Pair_Eff}).

\subsection{Vertical structure}
 \label{ssec:VertStru}

	In the absence of detailed knowledge on the `vertical'
 deposition of shear energy, little can be said about the thermal gradient
 along $\unvc_\theta$.  The entropy gradient is thus assumed to be 
 convectively stable.  Strong stratification, as it occurs in neutron stars
 where the field has to reach super-equipartition values to break free 
 buoyantly (Thompson 1994), is unlikely to occur in a disk where 
 self-gravity is negligible.

	Narayan and Yi (1995a) have pointed out that, 
 at least in the self-similar regime, 
 ADAF's are well described by ``vertically" integrated versions of
 the fundamental equations as long as the boundary surfaces are spherical
 in flat geometry.  The averaging prescription of Eq [\ref{eq:VertInt}]
 (note a typo in GP98), generalizes this procedure to curved 
 three-space.  Because of curvature effects,
 $\cH_\theta 	\equiv r \Theta_\cH 
		\neq \int \! \sqrt{g_{\theta\theta}} \, d\theta$, i.e. 
 the pressure scale-height as calculated in the LRF does {\it not} correspond
 to a locally integrated meridional line element.  $\cH_\theta$ is merely a 
 convenient definition of ``height" using the BLF `$r$' coordinate.  
 This definition breaks down when the disk is very thick, $\Theta_\cH > 1$,
 but for $\Theta_\cH \ltaprx .4$ 
 the correction is small ($\approx 2.5\%$ for $a = .95$).  
 The instability lengthscales in \S \ref{ssec:t_l-scales} should 
 correspond to locally integrated line elements but, 
 ignoring this small correction, the pressure
 scale height is used as the standard normalization.  

	Abramowicz, Lanza and Percival (1997, hereafter ALP)
 have derived an equation for the pressure scale height 
 which contains only thermodynamic potentials,
 geodesic flow invariants, and the meridional 4-velocity component and its
 radial gradient.  In the present situation, buoyant instabilities may render
 the velocity terms non-negligible, but one may still expect the fluid to be
 subsonic or transonic far from the horizon; effectively ``freezing out" close
 to the horizon (i.e. flowing along $\theta \simeq$ const.).  In this case, 
 ALP give
\beq
	\cH_\theta^2 	\simeq 2 ~ {p_0 \over \varrho_0} \, {r^4 \over \cL}
	 		= {2\over \Gamma} \, {c_{s}^2 \over {\Omega_z^2}}
\label{eq:H_def} \eeq
 where $\cL = \ell^2 - a^2(\sE - 1)$, $c_s$ is the effective soundspeed
 at the midplane of the disk and $\Omega_z$ is an effective frequency of 
 meridional oscillatory motion (GP98).

\section{An Instability-driven Dynamo}
 \label{sec:IDD}

	In spite of  being the subject of great theoretical effort since 
 its inception (Sakura \& Sunyaev 1973), clear resolution on the 
 question of the origin of anomalous viscosity in accretion disks 
 (Pringle 1981) has only recently occurred.   Hawley, Balbus and Winters 
 (1999) conclude that non-magnetic hydrodynamical Keplerian accretion is
 stable (see also Godon and Livio 1999) and that shear-fed MHD
 turbulence in a fully developed state is capable of explaining 
 anomalous angular momentum transport in accretion disks.
 (Balbus and Hawley 1991 (BH91), 1992a (BH92a), 1992b, 1996, 1998; 
 Hawley and Balbus 1991, 1996; Hawley, Gammie and Balbus 1995;
 Brandenburg, Nordlund, Stein \& Torkelsson 1995, 1996; 
 Stone, Hawley, Gammie \& Balbus 1996; Brandenburg 1998).

 	MRI induced magnetohydrodynamic turbulence may thus be reasonably
 proclaimed to be {\it the} source of anomalous angular momentum transport
 in accretion disk systems and the question of whether large scale fields
 are present must be subjected to their co-existence with the turbulent
 flow.

	Because the IDD is shear-fed, where the disk is advective and gas
 pressure dominated the low shear rate of an ADAF composed of non-relativistic 
 particles gas starves the dynamo out (recall that when
 $\Gamma_{\rm effective} \rightarrow 5/3$, self-similar solutions have
 nearly spherical accretion $\Omega \rightarrow 0$, Narayan \& Yi 1994, 
 Narayan, Mahadevan \& Quataert 1998).
 We thus consider the case where a large-scale field
 in the outer disk begins to emerge where its influence is dynamically
 unimportant (e.g., at $r \approx 70$, once photodisintegration is complete).

\subsection{On the effect of a radial flow}
 \label{ssec:RadFlow}

 	Although a predominantly toroidal field may be expected {\it a priori}
 (at least on surfaces perpendicular to $\unvc_\theta$), the radial flow 
 is non-negligible in the innermost disk.  For instance,
 in the standard model of PWF, the ratio 
 $\Re_{\varphi/r} \equiv (\gamma_\varphi^{-1} \beta_r) / \beta_\varphi$
 which compares the radial and azimuthal three-velocity components as 
 measured in the LNR frame, is small but significant in the range 
 $r \simeq [70,~4.5]$.
 At $r \simeq 14$ (where $\Theta_\cH \approx .4$ is maximum),
 the sub-Keplerian $\beta_\varphi \simeq .26$ and $\Re \simeq .12$. 
 At $r \simeq 4.5,~ \beta_\varphi \simeq .62$ and $\Re \simeq .32$ while 
 at $r \simeq 3.9,~ \beta_\varphi \simeq .61$ and $\Re \simeq .43$ and 
 quickly rising.  

	We envisioned that the MHD condition of high conductivity 
 in a strongly sheared semi-Keplerian flow produces a mean-field on 
 $\theta =$ constant surfaces that tracks the mean-flow on average 
 (equivalently, one may assume vanishing angular velocity gradients along 
 {\it mean} field lines).
 Thus, since the presence of a small radial field component
 does not alter the Balbus-Hawley instability (BH91), one may 
 cautiously  ignore the effect of a small, passively advected, 
 $B_r$ component also on the (linear) development of non-axisymmetric 
 $\cM$-type instabilities as calculated form a purely toroidal field 
 (\S \ref{ssec:NonAxi-MRI}).  Note, however, that these two
 MRI's originate from different branches of the MHD wave dispersion
 relation (the slow and \Alf branch).  This assumption could require
 re-evaluation when $B_r \gtaprx B_\varphi$.  
 
\subsection{On the use of spherical coordinates} 
 \label{ssec:SpheCoor}

	To date, most magnetohydrodynamical stability analyses of accretion
 disk systems have been performed in the context of the thin disk formulation 
 where the use of cylindrical polar coordinates is customary.  
 In a semi-thick, slim disk setting, the polar $z$ coordinate is preferably
 replaced by the meridional distance $\simeq r \vartheta$ 
 (\S \ref{ssec:VertStru}). 

 	A fundamental difference in the use of spherical coordinates
 is that while in a thin disk setting the vertical pressure gradient
 is balanced by the component of gravity along $\unvc_z$, 
 gravity plays no role in balancing pressure gradients along
 $\unvc_\theta$ (ALP). 
 Gravity's place is taken by the (inertial) centrifugal force
 which for rotation on spherical shells 
 is $\propto \sin^2 \! \vartheta$ (with $\vartheta = \pi/2 -  \theta$).
 In order to compare this with ``gravity along $\unvc_z$", one notes that 
 in the standard thin disk approach 
 $g_z$ is $\propto \sin \vartheta$ 
 and so the fractional difference is of 
 $\cO (\Theta_\cH) \approx  \sin \! \vartheta$.
 It follows that {\it the gradient 
 in the inertial force along $\unvc_\theta$ is gentler than that of 
 gravity along $\unvc_z$} 
(at least for $\vartheta < 1$)\footnote{
 A general relativistic analysis of this issue is complicated by the lack
 of a precise definition for the centrifugal force in a stationary metric
 (although such definition exists for a static (Schwarzschild) metric as 
 demonstrated through the formalism of the `optical reference geometry', 
 Abramowicz, Carter \& Lasota 1988, Abramowicz 1990, 
 Abramowicz \& Prasanna 1990).  Note, however,
 that the meridional hydrostatic equilibrium equation involves 
 an effective frequency of oscillation, Eq [\ref{eq:H_def}] (next footnote), 
 while the condition of radial force equilibrium, Eq [\ref{eq:SlimKerr}],
 involves deviations from Keplerian rotation rates instead
 (Appendix \S \ref{apx:RelMod2Dis}). 
}.  Note that although this somewhat diminishes the effects of buoyancy
 and the impact of the inertial force gradient (not gravity)
 on the Parker instability, for 
 $\cH_\theta/r \ll 1$ this effect is not severe.  

	Another difference is that the scale height in spherical coordinates 
 is curvilinear.  The scale height enters the general dispersion relation
 of the Balbus-Hawley (1991) instability by establishing a characteristic 
 lengthscale for the (predominant) vertical--gradient-dependent part of 
 the Brunt-V\"ais\"al\"a frequency (c.f. their Eq [2.8]).  This gradient 
 is approximately inversely proportional to the pressure scale height for
 adiabatic changes in the Boussinesq approximation.  In the `horizontal
 regime' $k_r/k_\theta \rightarrow 0$, however, this dependence drops out
 of the dispersion relation (BH92a) while simultaneously yielding the maximum
 growth rate.  The dispersion relation is simplified further by assuming
 rotation on cylindrical shells (i.e. coinciding isobaric and isochoric 
 surfaces).  An equivalent assumption in spherical coordinates for a slim
 disk is rotation on spherical shells and with this, BH91 analysis is 
 straightforwardly carried on to a thicker disk.  

	Non-axisymmetric $\cM$-type instabilities of a purely toroidal
 field only require (\S \ref{ssec:NonAxi-MRI}) that 
 4A$ \Omega + k_\varphi \! \Valf_\varphi < 0$ for 
 non-relativistic Keplerian rotation in the horizontal
 regime (Foglizzo \& Tagger 1994, hereafter FT94). 
 Although the geometry imposes a lower bound 
 on $k_\varphi = 1/r$ for nearly Keplerian rotation, the above restriction is
 {\it independent} of scale height and constraints only the toroidal field
 strength.  To first order, general relativistic corrections 
 (\S \ref{apx:RelMod2Dis}) do not modify this conclusion.  
 Indeed, magnetorotational instabilities in the horizontal regime
 are generally insensitive to vertical gradients and the pressure scale 
 height enters only indirectly when the \Alf speed is scaled with 
 respect to the local sound speed of the medium 
 (see the discussion at the end of \S \ref{ssec:t_l-scales}). 

	Lastly, since buoyant motions are commensurate on stratification
 and since radial pressure gradients force rotation on nearly spherical
 shells when the shear is strong (i.e. in a mildly advective disk), 
 the B$_\perp$ flux should naturally rise along the $\unvc_\theta$
 local pressure gradient.   This is a crucial point that suggests 
 that a fraction of the magnetic flux generated by the dynamo is 
 promptly collimated into the funnels above and below the disk's 
 surface thus creating favorable conditions 
 for the formation of a Poynting jet.

\subsection{The dynamo equations}
 \label{ssec:DynEqu}

 	The shear flow generates $B_{{\varphi}}$ 
 from $B_{{r}}$ on the shear timescale $\tau_{_\cS}$.   
 In the absence of a propitious field topology for the
 development of Parker's undulate instability (e.g. FT94), 
 the buoyancy of the B$_\perp$ flux, which escapes
 the disk on a timescale $\tau_{_\cB}$, limits the growth
 of $B_{{\varphi}}$ and $B_{{r}}$ and {\it may} generate 
 $B_\theta$ albeit inefficiently:  
 The flux buoyancy acts as an interchange instability whereby 
 inhomogeneities along a field domain cross section induce a 
 gradient in the buoyant velocity along the direction of the 
 mean field thus generating a small $B_\theta$ component.
 Inclusion of this rather uncertain process for meridional 
 field generation adds considerable mathematical complexity to
 the heuristic equations and will be overlooked in the name of simplicity.
 Note, however, that although the meridional field is unlikely to attain
 strengths that could explicitly alter the proposed dynamo equations, 
 its presence does alter the equilibrium turbulent state since the poloidal 
 MRI is more efficient at feeding the turbulence (E. Vishniac, Priv. Comm.).  
 This affects the IDD explicitly through our estimate of the magnetic
 Mach number for the turbulence (\S \ref{ssec:Turb_Rec}).

 	Non-axisymmetric $\cM$-type instabilities that develop
 in a purely toroidal field (FT94, Foglizzo \& Tagger 1995 [FT95], Terquem 
 and Papaloizou 1996 [TP96], Papaloizou and Terquem 1997) contribute
 to both $B_{{\varphi}}$ and $B_{{r}}$ field generation but their growth 
 rates decrease with field strength, c.f. Eq [\ref{eq:M_Tscal}],
 and are likely to be important only
 for $B_{{r}}$ since shear is the primary process for azimuthal field 
 growth.  TP96 note that because the toroidal MRI  
 is essentially local in azimuth (at least for small fields)
 the field need not be entirely toroidal, only {\it local toroidality}
 is required.  

	Additionally, reconnection of  all field components
 proceeds at a rate that is proportional to the upstream component of the 
 \Alf velocity, {\bf v}$_{\rm Alf}$ = {\bf B}$/\sqrt{4\pi\varrho}$, times 
 the square of the {\it magnetic} Mach number of the turbulence 
 (Lazarian \& Vishniac 1999) as measured in the LRF.
 
	In Lagrangian, LRF coordinates, the phenomenological 
 two dimensional dynamo equations are self-evident
\begin{eqnarray}
 \partial_t B_\varphi 	&=& {1 \over \tau_{_\cS}} 	B_r
			- {1 \over \tau_{_\cB}}		B_\varphi
			- {1 \over \tau_r} 	B_\varphi 	\cr
&&\cr
 \partial_t B_r		&=&  {1 \over \tau_{_\cM}} 	B_\varphi
			- {1 \over \tau_{_\cB}}		B_r
			- {1 \over \tau_r} 		B_r	
\label{eq:EurDynEqs}
\end{eqnarray}
 Within the heuristic phenomenology of these equations, when competing 
 gain (loss) timescales differ greatly one may safely drop the weaker
 terms.

\subsection{Relativistic scalings}
 \label{ssec:t_l-scales}

 	In a classical Keplerian disk, the timescales related to
 shear, magnetorotational and local {\it vertical} oscillations of test 
 particles are all scaled by the local Keplerian rotation frequency (as 
 meassured by observers at rest in a global frame).   In a relativistic 
 slim disk setting, however, the timescales related to all of these processes 
 are slightly different.   The importance of this remark becomes evident 
 when one attempts to generalize the instability scalings (LRF time and 
 length scales) for a slim disk in the Kerr-Newman geometry.

	Through Oort's A parameter (see below), the shear and 
 magnetorotational timescales are intrinsically related to the azimuthal
 rotation frequency of the disk.  For nearly Keplerian
 prograde rotation the natural scaling is simply $\Omega_+$ 
 (c.f. Eqs [\ref{eq:ArelKep}] \& [\ref{eq:BH_Tscal}]).  

	On the other hand, the natural frequency of local
 {\it meridional} oscillations of the four
velocity GP98\footnote{
 obtained by expanding Carter's 
 (1968) fourth constant of geodesic motion in the Kerr geometry, $Q$, 
 about $\vartheta \equiv \pi/2 - \theta$ and taking an affine derivative 
\[	
 	U^{-1}_\vartheta d_\tau \left\{
	Q = 	U_\theta^2 + 
	\cos^2\theta \left({l^2\over{\sin^2\theta}} - a^2 (\sE^2 - 1)\right)
	\right\} 
\]}, $\Omega_z$,
 which is directly involved in determining the local pressure scale height
 $\cH_\theta$, c.f. Eq [\ref{eq:H_def}], differs from the prograde Keplerian
 rotation rate by a factor $\Re$, where (Riffert \& Herold 1995)
\beq
 \Re^2 	\equiv \left( {\Omega_z \over {\Omega_+}} \right)^{\!\!2}
	\simeq \frac{1 - 4a r^{-3/2} + 3a^2 r^{-2}}
		{(1 + a r^{-3/2})^2 (1 - 3 r^{-1} + 2a r^{-3/2})}.
\eeq

	As advertised, the growth rates of $\cM$-type instabilities are
 characterized by a relativistic generalization of Oort's A parameter 
 while for buoyant motions and reconnection timescales 
 (and $\cP$-type instabilities) the characteristic timescale is
 (loosely speaking) the meridional `\Alf transit time': 
 ${\cH_\Theta / \Valf_{\perp\theta}}$.
 Thus, although the pressure scale height, $\cH_\Theta$, naturally 
 normalizes all meridional lengths,
 similarly scaled magnetorotational wavenumbers must be corrected by 
 a factor $\Re$ from their classical counterparts
\beq	k^{\rm Rel}_{_\cM} \longrightarrow \Re \, k^{\rm Class}_{_\cM}.
\label{eq:WvMumFix} \eeq

 	That this is indeed the case may be more easily seen by normalizing 
 either the classical (with $\Omega \equiv \dot{x}^\varphi/\dot{x}^t$)
\[
	4 {\rm A}_\Oor \Omega + ({\rm \bf k \cdot v}_{\rm Alf})^2 < 0
~~~~~~~~~~~~~~~~~~~~~~~~~~~~~~~~~~~~~~~~~~~~~~~~~~~(\ref{eq:ClasStab})
\]
 or the generalized relativistic stability criterium for shearing
 instabilities, Eq [\ref{eq:RelInStab}].  Because for prograde 
 Keplerian orbits the term driving the radial differential 
 force scales as $\Omega_+^2$, normalization of the \Alf velocity 
 to the local sound speed in the $({\rm \bf k \cdot v}_{\rm Alf})^2$ 
 term yields the correction factor if one desires to normalize the 
 wavenumber to the disk's pressure scale height.

\subsection{Linear shear}
 \label{ssec:lin_Sh}

 	We choose to model the shear flow in a linear shearing sheet
 approximation (Goldreich \& Lynden-Bell 1965) noting the well known
 limitations of such an approach, i.e. ignoring global curvature 
 (Ogilvie and Pringle 1996, TP96) and non-linear shear effects.  
 The influence of shear is thus parameterized by a relativistic 
 generalization of Oort's first constant, $\tau_{_\cS}^{-1} = $
 2A$_\Oor^{\rm Class} = d_{\ln r} \Omega~~ (= \3ov2 \, \Omega_\Kep$,
 for a Keplerian disk).

 	A relativistic expression for Oort's A value that contains
 the radial gradient of $\ell$ and $\beta_r$ implicitly is 
 (Appendix \S \ref{apx:RelMod2Dis}) 
\[
 	\sigma_{\hat{r}\hat{\varphi}} (r, \,d_r \Omega) 
		= \sA \gamma^2    
		\left\{\half  {d\Omega \over {d \ln r}} \right\} 
		\equiv {\rm A}^{\rm Rel}_\Oor. 
~~~~~~~~~~~~~~~~~~~~~(\ref{eq:OortsA}) 
\]
 This result differs subtly from the one in Novikov and Thorne 
 (1974) where $\gamma$ goes to $\gamma_\varphi$ for a thin disk 
 with negligible radial flow. 

 	For nearly prograde/retrograde Keplerian orbits, 
 from Eq [\ref{eq:OortsA}] one finds 
\beq
	\left| { 1 \over {\tau_{\cS}^{\rm Rel} }} \right| \simeq
	{3\over2} \, \gamma^2 \sA \; {\Omega_\pm \over {1 \pm a r^{-3/2}} }
	\equiv -2{\rm A}^{\rm Rel}_\Oor. 
\label{eq:ArelKep}\eeq
 Shear-fed instabilities, on the other hand, pertain to growth rates
 slightly different from A$^{\rm Rel}_\Oor$ due to relativistic 
effects\footnote{ 
 The relativistic shearing and magnetorotational  
 timescales are derived by modifying only the relativistic analog to the 
 classical ``radial differential force" (FT94), and {\it assuming} that,
 both, the centrifugal and the coriolis forces act as they do classically, 
 c.f.  Eqs [\ref{eq:Anis_Hill}] and footnote and discussion at the end of 
 Appendix \S \ref{apx:RelMod2Dis}.
}.

	Shear also forces the radial wavenumber
 of perturbations to evolve according to 
\beq 	k_r(t) = k^0_r - 2{\rm A} k_\varphi t.
\label{eq:RaWaNuEv}\eeq
 However, we shall concern ourselves 
 only with the horizontal regime
 of magnetorotational instabilities, $k_r/k_\theta \rightarrow 0$, 
 noting that this constraint yields the highest growth rate from both the 
 Balbus-Hawley (BH92a) instability (\S \ref{ssec:Pol_BH}) and toroidal, 
 non-axisymmetric $\cM$-modes (\S \ref{ssec:NonAxi-MRI}).

\subsection{The poloidal (Balbus-Hawley) MRI}
 \label{ssec:Pol_BH}

 	Classically, the axisymmetric magnetorotational
 instability (Balbus \& Hawley 1991) only requires that $\partial_r \Omega < 0$
 and that the $B_\theta$ field be ``weak" in the sense 
 $\Valf_\theta < (1/\pi) \, |A\Omega| \, \cH$ which is derived by imposing
 an upper limit ($2\cH$) on the `vertical' wavelength in the Keplerian thin 
 disk limit (note that the radial field component is irrelevant in the
 horizontal regime: $k_r/k_\theta \rightarrow 0$ BH92a). 

	On the other hand, the classical instability criterium 
 (Chandrasekhar 1961 and Appendix \S \ref{apx:RelMod2Dis}), 
 $4{\rm A} \Omega + k_\theta \Valf_\theta < 0$, sets a upper limit on 
 $k^{\rm max}_\theta = \sqrt{3} \, \Omega / \Valf_\theta$ beyond which the
 energy cost of bending field lines on small scales suppresses the instability.
 The fastest growth occurs for $k_\theta = \Omega/\Valf_\theta$ 
 at a rate that equals A$^{\rm Class}_\Oor$ as long as
 $k_\theta \gtaprx \pi/\cH_\Theta~~(=\pi\sqrt{\Gamma/2}\times\Omega/c_s)$ 

 	Although a fully general relativistic stability analysis of MHD
 disk systems in the Kerr geometry is beyond the scope of this paper, 
 in Appendix \S \ref{apx:RelMod2Dis} we derive an approximate (see
 footnote 6) relativistic timescale, corrected for the form of the 
 ``radial differential force" in the Kerr geometry.  
 For nearly Keplerian orbits,
 this procedure leads to replacement of the term driving the 
 (classical) radial differential force $d_{\ln r} \Omega^2$ 
 by $4 \Omega_\pm {\rm A}^{\rm eff}_{\cM\pm}$ 
 For prograde orbits; using 
 Eqs [\ref{eq:OortsA} \& \ref{eq:RelMRRate}] 
 to estimate the timescale and
 Eq [\ref{eq:WvMumFix}] to normalize the optimal wavenumber, 
 one finds
\beq
	\left| { 1 \over {\tau_{_\BH}^{\rm Rel} }} \right|_{\rm max} 
	\!\!\!\!=
	{3\over4} \, \gamma^2 \left[ { \sA \over {\sD}} \right] \Omega_+
	~~~\left(\equiv -{\rm A}^{\rm eff}_{_\cM +} 
		~\right)
\label{eq:BH_Tscal} \eeq
 (compare this with Eq [\ref{eq:ArelKep}]),  and
\beq
	\cH_\Theta \, k_{_\BH} 	= \sqrt{2 \over \Gamma} {c_s \over \Omega_z} 
			\, \Re \, {\Omega_+ \over \Valf_\theta}
			= \sqrt{2 \over \Gamma} \, {c_s \over \Valf_\theta}.
\label{eq:BH_Kscal} \eeq
 We identified the r.h.s of Eq [\ref{eq:BH_Tscal}] with the 
 {\it maximum} growth rate of shear-fed instabilities (such as $\cM$-type) 
 in relativistic prograde Keplerian Kerr disks as meassured in the LRF 
 of the fluid:
\[ 	\left| \frac {1} { \tau^{\rm Rel}_{_\cM}} \right|_{\rm max, +}
	\!\!\!\!\!\!\!\!\!\!  =  
	-{\rm A}^{\rm eff}_{_{\cM+}}  
\] 

\subsection{The non-axisymmetric toroidal MRI}
 \label{ssec:NonAxi-MRI}

	Although BH92b originally found that the most unstable 
 magnetorotational modes correspond to axisymmetric modes evolving in
 a poloidal field configuration (through destabilization of \Alf modes), 
 several authors (FT94, TP96, Matsumoto \& Tajima 1995) find similar
 growth rates for non-axisymmetric modes in a purely toroidal field
 configuration (albeit at larger poloidal wavenumbers, Balbus 1998).  
 Note that in hyper-accreting 
 \gam-ray burst black holes, non-axisymmetric modes may be 
 strongly excited if these feed the angular momentum loss
 through an associated gravitational wave emission.  

 	This instability corresponds to destabilization of the slow MHD mode
 of wave propagation (FT94, TP96) as is also the case for Parker's undulate 
 instability.  FT94 (see also TP96, Shu 1974)
 remark that the inertial effect of rotation enters through the coriolis
 force term, slightly twisting flux tubes of displaced fluid elements 
 in the azimuthal direction, and through the radial differential force.

  	If the ratio $k_r/k_\theta \ll 1$~ ($\Rightarrow$ `horizontal' 
 regime), meridionally localized disturbances are more prone to the radial
 differential force.  If this force is strong enough to overcome the magnetic
 tension this results in the non-axisymmetric magnetorotational instability 
 (FT95).  Radially localized perturbations $k_\theta/k_r \ll 1 ~~(\Rightarrow$ 
 `asymptotic' limit, c.f. Eq [\ref{eq:RaWaNuEv}]), on the other hand, 
 would naturally evolve into Parker unstable modes if the field 
 topology were favorable.  
 More generally, this instability generates both $B_r$ and 
 $B_\varphi$ (in the horizontal regime, the slow mode is essentially
 an \Alf wave) and affects the field configuration by twisting of
 toroidal field lines into flux ropes perhaps enhancing
 turbulent pumping (Vishniac 1995a)
 and the associated field buoyancy (\S \ref{ssec:Pair_Eff}).
    
 	We use the Kerr geometry generalization of the radial differential
 force, \S \ref{apx:RelMod2Dis}, and define 
 $\alpha_\varphi \equiv P_{{\rm B}_\varphi} / (\Pgas + \Pb)$ 
 and a rescaled $\tau_\cM$ and $\AMp ~(< 0$) 
\beq 
	\hat{\tau}_{_\cM}  \equiv  { \tau_{_\cM} \over  \Omega_+},
~~~~~~~ 
 	\AM \equiv {\AMp \over  \Omega_+},
\label{eq:ReScaA} \eeq
 to adopt the results in FT95 as approximate
 relativistic generalizations of the toroidal MRI scales.  

 For prograde, nearly Keplerian rotation
\begin{eqnarray}
	| \AMp |^2 \tau^2_{_\cM}   &=&	
~~~~~~~~~~~~~~~~~~~~~~~~~~~~~~~~~~~~~~~~~~~~~~~~
\cr &&\cr
 && \!\!\!\!\!\!\!\!\!\!\!\!\!\!\!\!\!\!\!\!\!\!
    \!\!\!\!\!\!\!\!\!\!\!\!\!\!\!\!\!\!\!\!\!\!
 	{1 \over 2} \left\{ 
	\sqrt{(1+2\alpha_\varphi)(1+2\alpha_\varphi(1+\AM))} 
	+ 1 + \alpha_\varphi (2 + \AM) 	\right\}	
\label{eq:M_Tscal} \end{eqnarray}
 and
\begin{eqnarray}
 \cH^2_\Theta \, k^2_{_\cM}  &=& 
		-{2 \over \Gamma} 
		\left( {c_s \over \Valf_\varphi} \right)^{\!\!2} \times 
 		( 2 \AM +  (1 + \alpha_\varphi) \hat{\tau}_{_\cM}^{-2} ) \cr
&&\cr
		&\equiv& 
		-{2 \over \Gamma} 
		\left( {c_s \over \Valf_\varphi} \right)^{\!\!2} 
		\times ~~~\hat{\delta} 
\label{eq:M_Kscal} \end{eqnarray}
 for the non-axisymmetric MRI modes of most vigorous growth.

 	These estimates are in fact subject to the condition 
\beq 	-\AM < 1 + (2\alpha+1)^{-1}.
\label{eq:trouble} \eeq 
 In a stronger {\it effective} shear flow the optimal wavenumber in
 Eq [\ref{eq:M_Kscal}] becomes imaginary and the toroidal field is 
 unstable to a faster radial interchange ($k_\varphi = 0$) instability  
 according to the Rayleigh criterion (T. Foglizzo, Priv. Comm.).  
 Interestingly, this destabilization process is mathematically
 similar to the one that occurs when suprathermal particles 
 push the Parker instability into interchange-like mode
 (FT94) for $p_{\rm suprathermal}/\Pgas > 3 + 4\alpha$ (see below).

 	To the IDD, this means that when condition [\ref{eq:trouble}] is not
 meet, the toroidal MRI does not grow the radial field nor does it provide
 structure nor field reversals in the linear analysis.  Moreover, note 
 that when the wavenumber becomes very small, the local analysis is out
 of its intended regime and global curvature effects become important.
 The IDD is thus inadequate in this limit.
 Fortunately, condition [\ref{eq:trouble}]
 is only meet at most within $r \ltaprx 5 ~[GM/c^2]$ for a non-rotating
 black hole and it moves inward for a rotating black hole 
 (\S \ref{ssec:results}).

\subsection{On the buoyancy induced by \\ radiation and $e^\pm$ pairs}
 \label{ssec:Pair_Eff}

	One may physically expect that $e^\pm$ pairs would enhance the 
 buoyancy of magnetic flux if such a lighter fluid component coupled to
 the `other' light relativistic fluid of the medium, i.e. the magnetic field. 
 Yet, further elaboration on such a conjecture requires some understanding
 of the dynamics of quasi-equilibrium pair plasmas in 
 magnetic fields well in excess of the Q.E.D. field scale
 ($B_{\rm Q} = m^2/e (c^3/\hbar) \simeq 44.$ TG),
 a problem that has yet to be explored.

 	For instance, under this circumstance pairs are created (annihilated)
  mainly {\it via}~ $1\gamma$ decay(production) with strongly asymmetrical 
 energy profiles for each member of the pair (Daugherthy \& Harding 1983, 
 Harding 1986).  
 It is thus conceivable that at very high disk temperatures the pairs 
 may have considerable energy density in a non-thermal distribution
 of suprathermal particles.  This being the case, Parker's suprathermal 
 mode of hydromagnetic wave propagation (which strictly speaking applies
 in the collisionless regime) will enhance Parker's undulate instability 
 possibly pushing the latter into an exchange-like mode
 (FT94) with a normalized growth rate
 $(\omega_{\rm Growth}/\Omega)^2 
 	\propto (\alpha_\perp-\beta)/(1+2\alpha_\perp)$,
 where $\alpha_\perp$ is calculated from the B$_\perp$ field,
 i.e. including $B_r$ in contrast to 
 Eqs [\ref{eq:M_Tscal} \& \ref{eq:M_Kscal}].

	On the other hand, (thermal) transrelativistic pairs are well 
 coupled to the nucleon component even when coulomb dragg is low  
 (due to the neutronization of the material) and ambipolar diffusion 
 of the field+pair component through neutron matter is dismal 
 (although the turbulence may help the latter).  Thus, the most 
 likely mechanism for fast horizontal flux escape would have to 
 involve nucleon density deficits inside flux ropes or, in the case
 of a diffuse field, turbulent diffusion.  

	A plausible mechanism that promotes baryon unloading from field 
 lines is turbulent pumping (Vishniac 1995a) by the MRI which must
 favorably compete with turbulent diffusion of matter back onto flux
 ropes.  Under this assumption, the stretch, twist and fold of field lines
 by (enthalpy-weighted) sub-Alfvenic turbulence augments the field 
 energy density and releases matter from field lines that would
 otherwise be ``frozen-in".  For the marginal case of Alfvenic 
turbulence\footnote{ 
 Note, however, that numerical experiments bias our expectation of MRI 
 turbulence toward super-Alfvenic values, i.e. the Maxwell stress
 tends to dominate over the Reynolds stress.
},
 nearly empty B$_\perp$ flux ropes in a gas pressure
 dominated disk acquire a drag limited buoyant velocity 
 $v_b \propto (\Valf_{\perp\theta})^2/c_s$. 
 Moreover, assuming efficient diffusion of radiation and
 $e^\pm$ pairs into the flux tubes, in this picture 
 the buoyancy loss rate, $v_b / \cH_\theta$, is
 enhanced by a factor $\ltaprx ~ p/\Pgas$ (Vishniac 1995b). 
 
	On the other hand, in a diffuse field configuration the local
 value of the turbulent helicity implies migration of the field to flux
 poor regions of the accretion disk (E. Vishniac, Priv. Comm.).  
 Phenomenologically, this is also born in numerical simulations with 
 vertical stratification (Brandenburg, \etal 1995, Stone, \etal 1996).  
 Since the main contribution to the field generation occurs in regions 
 of highest pressure, i.e. the disk midplane, 
 this argument yields a systematic meridional
 motion of the field (up to a gradient in the diffusion coefficient)
\beq
 v_b 	= {2 \over \Gamma} \, {v_{\rm turb}^2 \tau_{\rm cor} \over \cH_\Theta}
	\simeq \sqrt{2 \over \Gamma} \Re \,
	 {(\Valf_{\perp\theta} )^2 \over c_s}  
 	\left\{ \tau_{\rm cor} \over \Omega^{-1}_+ \right\} 
	{\rm M}_{\bf B}^{t\, 2}
\label{eq:buoyV} \eeq
 where $\tau_{\rm cor} \simeq \tau_\cM$ is the correlation timescale 
 of the turbulence and M$_{\bf B}^t \equiv v^{\rm turb}/\Valf$~
 is its magnetic Mach number on the largest eddy scale.
  
 	The estimate for the field loss rate from the disk depends 
 on whether or not the radiation/pairs enhance the buoyancy loss
 rate.  In the context of a diffuse field configuration (which 
 is on more solid grounds), this enhancement is only a conjecture. 
 With this in mind, using Eq [\ref{eq:buoyV}] 
 and defining $\xi \equiv \Pgas/p$, 
 we write the characteristic time as a meridional \Alf `transit time'  
 (admittedly a misleading nomenclature given that 
 \Alf waves move along field lines)
\begin{equation}
	\tau_{_\cB} (B_\varphi, B_r) = {\cH_\theta \over v_b}  
		= \eta \, {\cH_\theta \over \Valf_{\perp\theta}}
		= \eta' \, \left( {c_s \over \Valf_{\perp\theta}} \right) \; 	
		{1 \over {\Omega_+}}
\label{eq:Buoy_Tscal} \end{equation}
  where 
\beq	
	\eta' 	= \sqrt{2 \over \Gamma} \, {1 \over \Re} \, \eta 
		= \xi \, \Re^{-2} 
	\left( {c_s \over \Valf_{\perp\theta}} \right) \,   
 	\left\{ \Omega^{-1}_+ \over \tau_{\rm cor} \right\} 
	\, {\rm M}_{\bf B}^{t -2}
\eeq

\subsection{Turbulent reconnection}
 \label{ssec:Turb_Rec}

 	It is well known that in spite of generally predicted low reconnection
 rates (Sweet 1958, Parker 1979), well studied systems such as the solar
 corona and chromosphere indicate that reconnection is fast once it
 begins (Dere 1996, Innes \etal 1997), essentially occurring at an
 order of magnitude below the \Alf speed.

	Following this observation, Lazarian \& Vishniac (1999) 
 show that in the Goldreich \& Sridhar (1997) model of strong MHD 
 turbulence, reconnection in a weakly stochastic field occurs at
 a fraction of the \Alf speed which equals the square of the
 magnetic Mach number of the turbulence M$_t^{\bf B}$.
 In a medium where MRI's feed the turbulence, 
 an {\it educated guess} of this number is 
 (Vishniac \& Diamond 1992, Zhang, Diamond \& Vishniac 1994)
\[ {\rm M}_t^{\bf B} \in [(\Valf/c_s)^{1/3},~~ 1.].
\] 
 We adopt the lower value noting that the Balbus Hawley instability
 is likely to be more efficient at feeding the turbulence and that 
 numerical simulations generally show a dominant Maxwell stress
 regardless of the initial seed field.

 	In keeping with our informal nomenclature, the reconnection
 rates are written as inverse \Alf transit times calculated from the
 component of the field that undergoes reconnection 
\[ 	\tau_{\rm rec} \approx {l^{\perp i}_{\rm rec} \over \Valf_i}
		= 
		\sqrt{2 \over \Gamma} \, {1 \over \Re} \, 
		{l^{\perp i}_{\rm rec} \over \cH_\Theta} \,
		{c_s \over \Valf_i} \; \Omega^{-1}_+.
\] 
 $l^{\perp i}_{\rm rec}$ is a 
 perpendicular lengthscale associated with the mean distance for reversal
 of the field component undergoing reconnection.  Let us label
 $l^\perp_{\rm rec}$ according to 
\[
 \delta {x}^r \rightarrow X,~~
 \delta {x}^\varphi \rightarrow Y,~~{\rm and}~~
 \delta {x}^\theta \rightarrow Z,
\]
 with a subscript denoting the component 
 of the field undergoing reconnection.

 	The perpendicular lengthscales associated 
 with the mean distance for field reversal are derived from 
 the fundamental linear lengthscales for coherent field pumping.
 The latter are supplied by half of the toroidal MRI's (wave)length
 scale (recall that in the horizontal regime one deals essentially with 
 \Alf wave destabilization).

	Azimuthal, radial field reversal, $Y_r$, directly involves the
 optimal wavenumber of the non-axisymmetric MRI (\S \ref{ssec:NonAxi-MRI}).
 This lengthscale is 
 sheared in the azimuthal direction thus coupling it to the radial, 
 azimuthal field reversal lengthscale, $X_\varphi$.  Note that 
 the meridional lengthscales associate with the poloidal MRI
 are not relevant since the meridional field is assumed to be 
 weak (albeit quickly unstable).
 
 	It has been argued (TP92) that the shear process generally 
 reduces radial lengthscales until these supply the fundamental 
 reconnection channel.  This is indeed the case when any of the 
 coherent field pumping process is slow when compared to shear, 
 $\tau \gg \tau_{_\cS} \simeq \Omega_+^{-1}$; but this is never the 
 case when the MRI is the culprit (even for the slower manifestations
 of the instability).   One may thus considerably simplify the 
 calculation by ignoring this aspect of shear in the reconnection
 process. 

 	On the other hand, following TP92, we obtain an estimate 
 of the radial, azimuthal field reversal lengthscale, $X_\varphi$
 by noting that the time evolution of wavenumbers implied by 
 Eq [\ref{eq:RaWaNuEv}] during one MRI timescale
 couples the azimuthal lengthscale to the radial lengthscale, 
 i.e. $l_y^\perp = (\tau_{_\cM}/\tau_{_\cS}) \times  l_x^\perp$.

 Thus
\beq	Y_r = {\pi \over {k_{_\cM}}} ~~~{\rm and}~~~
	X_\varphi \equiv \left( {\tau_{_\cS} \over \tau_{_\cM}} \right)
			\times {\pi \over {k_{_\cM}}}.	
\label{eq:YxXyRecSca} \eeq

\subsection{Equilibrium Solutions} 
 \label{ssec:Equil_Sol} 

 	In general, Eqs [\ref{eq:EurDynEqs}] are implicitly complex due to
 the non-linearities introduced by the field dependence of gain timescales, 
 Eqs [\ref{eq:BH_Tscal}, \ref{eq:M_Tscal}],
 the buoyancy loss rate Eq [\ref{eq:Buoy_Tscal}], and 
 the reconnection times which involve the perpendicular lengthscales 
 Eqs [\ref{eq:YxXyRecSca}].
 
	Scaling wavenumbers to the inverse pressure scale height 
 $k \rightarrow \hat{k} / \cH_\Theta$, 
 and defining 
\beq 	\sqrt{\Gamma \over 2} \, \varepsilon'_{r} \equiv  
	{Y_r \over \cH_\Theta} 
	=  	{\pi \over {\hat{k}_{_\cM}}} 
 ~~{\rm and}~~
 	\sqrt{\Gamma \over 2} \, \varepsilon'_{\varphi} \equiv 
		{X_{\varphi} \over \cH_\Theta}
	 ~ = {\AM \over \AM'} \, {\pi \over {\hat{k}_{_\cM}}}
\eeq
 and 
\[	\AM  = {\tau^{-1}_{_{\cM}} \over \Omega_+}  ~~~~~{\rm and}~~~~~ 	
	2 \AM' =  {\tau^{-1}_{_{\cS}} \over \Omega_+} 
\]
 (c.f. Eq [\ref{eq:ReScaA}], although these are positive definite), 
 a set of normalized dynamo equations for prograde orbits follows  
\begin{eqnarray}
 \partial_{t'} B'_\varphi 	&=& 2\AM' 	B'_r
	- {1 \over \eta'} 	\, B'_\varphi \,B'_\perp 
	- {1 \over \varepsilon'_\varphi}	\, (B'_\varphi)^2 , 
\label{eq:Ana1} \\ &&\cr
 \partial_{t'} B'_r 		&=& \AM 	B'_\varphi
	- {1 \over \eta'}	\, B'_r \,	B'_\perp
	- {1 \over \varepsilon'_r} \, (B'_r)^2 ,
\label{eq:Ana2} \end{eqnarray}
 where the fields are in velocity units and normalized to the 
 soundspeed ($B' = B \times \sqrt{4\pi\varrho} \, c_s$) and the time
 is normalized to the inverse of the prograde Keplerian frequency 
 $t' = \Omega_+ t$.

	In a steady state, these equations must satisfy
\begin{eqnarray}
 B'_r 	&=& { B'_\varphi \over {2 \AM '} } \, [{\rm M}_t^{\bf B}]^2 \times  
	\left\{ {\Re^2 \over \xi \AM} \,  {B'_\perp}^{\!\!2} 
	+ {2\sqrt{2}\over\pi} \, \Re \, {\AM' \over \AM} \hat{\delta} 
 \right\} , 
\label{eq:Ana_Dyn1} \\ &&\cr
 B'_\varphi 	&=& { B'_r \over {\AM} } \,\, [{\rm M}_t^{\bf B}]^2 \times   
	\left\{ {\Re^2 \over \xi \AM}  {B'_\perp}^{\!\!2} 
	~+~ {\sqrt{2}\over\pi} \, \Re \, {B'_r \over B'_\varphi} \hat{\delta} 
 \right\} ,
\label{eq:Ana_Dyn2} \end{eqnarray}
 which we solve using a multidimensional Newton-Raphson method
 (e.g., Numerical Recipes in C, Press \etal 1988).

\section{A toy calculation}
 \label{sec:toying}

\subsection{Self-critique}

	The instability driven dynamo depends self-consistently 
 on the relativistic shear, $2\AM'$, and MRI, $\AM$, timescales,
 on the coherent field pumping lengthscale supplied by the MRI, 
 $\hat{\delta}$, and 
 on the magnetic Mach number of the turbulence, M$^t_{\bf B}$. 
 A phenomenological scaling for the buoyant velocity in the turbulent
 medium, Eq [\ref{eq:buoyV}],
 follows from the limiting cases of nearly empty flux tubes
 and turbulent diffusion of a spread out field (and by analogy
 to other buoyant instabilities such as the Parker instability).

 	An open issue is whether a radiation pressure dominated
 environment helps the buoyancy of the field and what this means 
 for the magnetic energy deposition rate, particularly in the
 context of a \gam-ray burst.  If the emerging magnetic flux (in ropes) 
 is relatively baryon-free, this helps its escape from the disk. 
 On the other hand, the meridional diffusion of B$_\perp$ flux
 in a semi-thick disk setting also tends to separate the field
 from the (heavier) baryon component, aiding in the escape of 
 flux+radiation and pairs. 

	In order to address the overall energetics and to gain a
 qualitative feel of the postulated hydromagnetic energy conversion
 process, 
 we have constructed a toy model which adopts the pressure
 ratios from the standard model of PWF and (non--self-consistently) 
 assesses the energy loss from the buoyancy of the field.  
 Evidently, a realistic model should account for the back-reaction
 of the field on the flow which in the context of an accretion disk 
 means primarily the angular transport implied by the 
 -$r \varphi$- component of the magnetic stress  
 (note that in the notation of the previous section, 
 $\alpha^{\rm mag}_\SS = B'_r B'_\varphi$).  
 In the absence of other significant sources of `anomalous' viscosity,
 a self-consistent hydromagnetic accretion disk model is possible in 
 principle when our equations are solved with a compatible set of 
 relativistic hydrodynamic equations 
 (e.g. GP98, Abramowicz \etal 1997) under the
 assumption that the angular momentum is transported by the largest 
 magnetic eddies. 

 	The radial flow (\S \ref{ssec:RadFlow}) 
 is passively accounted for by carrying out the analysis in the comoving
 frame and assuming that the linear timescale of the non-axisymmetric MRI
 faithfully reflects the growth rate of the radial field (an assumption
 that is not free of criticism).   
 On the less optimistic side, the radial interchange instability
 (\S \ref{ssec:NonAxi-MRI}) of the B$_\perp$ flux in a strong effective
 shear field shuts the dynamo off at the innermost section of the disk 
 where most energy could be extracted.  This instability is already 
 present when $1 < -\AM < 1 + (2\alpha+1)^{-1}$ (T. Foglizzo, Priv. Comm.) 
 but with a slower growth rate than the MRI.  Thus, even in the absence
 of substantial meridional flux (Park \& Vishniac 1994), radial buoyancy 
 may change the picture we have portrayed hereby (on the brighter side, 
 the radial interchange opposes the advection of the magnetic field).
 In this regard, note that condition [\ref{eq:trouble}] depends only 
 on the toroidal field strength when $k_\varphi \rightarrow 0$ 
 but we cap the latter at $k_\varphi = 1/\varpi$  

\subsection{The magnetic energy deposition rate}
 \label{ssec:results}

	The local magnetic energy flux (from buoyancy) is obtained from 
 the part of the time derivative of the energy density which is lost to 
 buoyant motions
\begin{eqnarray}
	\cH_\theta f_{\bf B} &=& \dot{u}_{\bf B} 
	= {1\over{4\pi}} 
		\left\{ B_r \dot{B_r}  + B_\varphi \dot{B_\varphi}
		\right\}						\cr
 	&&\cr
	&=& \left[ {c^3 \over {GM}} \right] \, r^{-3/2} \, 
		{{\Gamma \, p} \over \xi} \, {\Re^2 \over \AM} \,
		[{\rm M}_t^{\bf B}]^2 \times {B'_\perp}^{\!\!4} 
\label{eq:Flux_Dep} \end{eqnarray}
 where $\dot{B_r}$ and $\dot{B_\varphi}$ equal the second terms in 
 Eqs [\ref{eq:Ana1} \& \ref{eq:Ana2}].   	

 	Adopting the results from the standard model of PWF
 (c.f. \S\S ~\ref{sec:Intro} \& \ref{ssec:RadFlow}),
 this form for the specific power output is used in four estimates 
 of the steady magnetic flux deposition rate for $r \in [4.5, ~40]$:
 non-rotating black hole with (hBuoy-a.00), and without enhanced buoyancy
 (lBuoy-a.00), and nearly maximally rotating black hole, $a=.95$, 
 with (hBuoy-a.95), and without (lBuoy-a.95) enhanced buoyancy.  
 Curiously, because the fields are higher in the low buoyancy
 scenario, these yield the highest output rates 
 (in erg sec$^{-1}$): 
 $8.1_{+51}$ (lBuoy-a.00) and $1.0_{+52}$ (lBuoy-a.95) {\it vs} 
 $5.1_{+51}$ (hBuoy-a.00) and $6.6_{+51}$ (hBuoy-a.95).

 	The `half-luminosity' radius displays some interesting
 behavior as well: for hBuoy-a.00, the IDD becomes operational 
 (c.f. radial interchange instability)
 at $r_{\rm min} \simeq 5.3$ and 
 $r_{\cal L} \simeq 10.3$, while  			
 for lBuoy-a.00, $r_{\rm min} \simeq 5.9$ and 		
 $r_{\cal L} \simeq 9.2$,  				
 because of the higher stationary fields in the low buoyancy case:
 $[B'_\varphi,~B'_r] = 	 [.72,~.80]$ 			
 {\it vs}  		$[.42,~.48]$ at $r_{\rm min}$, 	
 respectively.

	For nearly maximally rotating holes, the IDD operates down to 
 smaller radii because frame dragging yields a slower effective shear:   
 $r_{\rm min} \simeq 5.4$, $B'_\varphi = .93$ and	
 $r_{\cal L} \simeq 8.5$ for lBuoy-a.95 and 
 $r_{\rm min} \simeq 4.6$,  $B'_\varphi = .57$ and 	
 $r_{\cal L} \simeq 8.5$ for hBuoy-a.95.

	The implied $\alpha_\SS$-parameter
 at the minimum operational radius is highest for 
 maximally rotating, low buoyancy disks $\approx .80$, 
 and smallest for non-rotating, highly buoyant disks $\approx .21$.
 This magnetic stress decreases quickly at larger radii:
 for low buoyancy,					
 $\alpha_\SS \approx$ .39, and .31 at $r =$ 10, and 20, respectively;
 while for high buoyancy,
 $\alpha_\SS \approx$ .16, and .17 at $r =$ 10, and 20. 

\subsection{Poynting jets and \gam-ray burst engines}

	We have constructed a reasonable set of heuristic two dimensional
 dynamo equations for magnetic field components in the comoving frame 
 under the premise of negligible generation of meridional field.
 These equations approximately account for field generation by the shear
 flow and by the non-axisymmetric MRI in the Kerr-Newman geometry.  
 Two local processes 
 are invoked as field loss terms: turbulent flux buoyancy (or vertical 
 diffusion) and turbulent reconnection.   Self-consistent equilibrium 
 solutions to these equations are found in proximity to 
 equipartition field values by adopting pressure ratios and
 the scale height from the standard model of PWF.  
 Although, one must not take these estimates at face value 
 until a self-consistent calculation is displayed, the total 
 hydromagnetic energy deposition rate from buoyancy is found to be 
 at least comparable to the neutrino luminosity in that scenario, 
 $L_\nu \simeq 3.3_{+51}$ erg sec$^{-1}$. 

	The magnetic
 field is removed from the disk interior and its ultimate fate depends on 
 the details of the disk corona dynamics which I do not address here. 
 At $\theta \sim \Theta_\cH$, some of this flux can be expected to undergo
 an inverse cascade to form larger coherent field structures (Tout and 
 Pringle 1995).  This large scale field is the relevant one with regards to
 Poynting jet production.  However, ignoring curvature effects, the field 
 generated by the non-axisymmetric MRI in a strong {\it effective} shear
 flow, c.f. \S \ref{ssec:NonAxi-MRI}, pushes the optimal 
 wavenumber to its lower 
 bound which implies pumping of coherent field structures with a 
 lengthscale of $\cO (\pi \varpi) \sim \pi r$.  In other words, the coherence 
 lengthscale of the toroidal MRI in a strongly sheared rotor is large.
 	Thus, in principle {\it all the energy deposited at the innermost
 radii may go into a Poynting flux}.  In the low buoyancy, high spin case,
 our estimate indicates that this may be as much as 10\% of the total 
 output rate, without invoking inverse cascade arguments.   Invoking 
 the Blandford-Znajek mechanism can only increase the above estimate.
 If fully self-consistent models yield similar results, this could render 
 the proposed hydromagnetic energy conversion mechanism a more 
 viable alternative to neutrino-burst driven models of 
 \gam-ray bursts.

\ldw
	\hfill {\it To Nicole Angela~ 9/24/1998-}
\ldw

	ACKNOWLEDGEMENTS: \\
{\it 	I am indebted to E. Vishniac for very stimulating discussions on 
 accretion disk magnetic field dynamics and MHD turbulent effects on dynamo
 processes; to T. Foglizzo for his thorough analysis and hindsight on the
 destabilization of the slow branch of MHD wave propagation; to D. Kazanas
 for his physical intuition approach to the problem at hand, to A.K. Harding
 for her constant unselfish help, and to N.A.S.A.'s ADS for making it possible
 for a large part of this research to be conducted while in my native 
 Costa Rica.  Most importantly, I am grateful to my wife and children 
 for their support. 	
 This investigation was initially supported by the University of Costa Rica 
 under grants \# 112-98-263 and \# 112-98-264.
}


\appendix

\section{Relativistic Shearing Instabilities}
 \label{apx:RelMod2Dis} 

	In a local kinematic analysis of the Hill (1878) equations for a
 ``sticky disk", Balbus \& Hawley (1992a) have shown that the equations 
 of motion for a (corotating) Lagrangian displacement vector $\bxi$ 
 (with $\unvc_r \rightarrow \unvc_x,$ and $\unvc_\varphi \rightarrow \unvc_y$)
\begin{eqnarray}
 	\xi_{,tt}^x - 2\Omega \xi^y_{,t} &=& 
		-\{4 {\rm A}_\Oor \Omega + \cK_x \} \xi^x \cr
&&\cr
 	\xi_{,tt}^y + 2\Omega \xi^x_{,t} &=& 
		~~~~~~~~~ - \cK_y ~~~~~~~ \xi^y,
\label{eq:Anis_Hill} \end{eqnarray}
 yield a dispersion relation identical to the one derived from an MHD 
 stability analysis of axisymmetric perturbations of a purely poloidal
 disk field configuration in the limit \mbox{$k_r/k_\theta \rightarrow 0,$}
 provided that 
\[ 	\cK_x = \cK_y = k^2_\theta v_{{\rm Alf},\theta}^2.
\]
 Notably, although both terms in the curly brackets of 
 Eqs [\ref{eq:Anis_Hill}] are stabilizing, their combined 
 influence is destabilizing (Balbus \& Hawley 1996) because
 of the spatial coupling.

	Foglizzo \& Tagger (1995) find an identical set of equations 
 for the magnetorotational instability of a purely {\it toroidal} field
 in the horizontal regime, provided that the ``spring constants" are 
 replaced by 
\[
 	\cK_x = k^2_\varphi v_{{\rm Alf},\varphi}^2 
	~~~~~{\rm and}~~~~~
 	\cK_y = k^2_\varphi \left( { {c^2_s v^2_{{\rm Alf},\varphi}} \over
	{c_s^2 + v^2_{{\rm Alf},\varphi}} } \right).
\]						
 This is not surprising given the generality of the
 magnetorotational destabilization process (BH92a).

	In either case, the dispersion relation is of the form 
 \mbox{$\tau^{-4}-b \tau^{-2}+c=0$}, with 
 \mbox{$b = 4(\Omega^2 + {\rm A}\Omega) + (\cK_x + \cK_y)$}, and 
 \mbox{$c = \cK_y (4 {\rm A} \Omega + \cK_x)$}, where we have set
 \mbox{$d_{\ln r} \Omega^2 = 4 {\rm A} \Omega$} 
 (recall \mbox{${\rm A}^{\rm Class}_\Oor = \half d_{\ln r} \Omega$}).
 This dispersion relation yields a generalized classical instability 
 criterium (i.e. $\tau^2 < 0$)
\beq
	4 {\rm A}_\Oor \Omega + ({\rm \bf k \cdot v}_{\rm Alf})^2 < 0
\label{eq:ClasStab} \eeq
 for magnetorotational instabilities in either poloidal or toroidal magnetic
 field topologies, in the horizontal regime $k_r/k_\theta \rightarrow 0$.
 Because in most cases of interest to us A$_\Oor < 0$, 
 the first term in Eq [\ref{eq:ClasStab}] 
 drives the destabilization process.

 	Classically, this term 
 (4A$\Omega$ \mbox{$ \equiv d_{\ln r}\Omega^2$}) may be 
 derived by considering the unbalance of inertial forces induced by deviations
 from Keplerian rotation of radially displaced fluid elements, i.e. induced by 
 the imbalance of centrifugal {\it vs.} radial gravity (FT95).  Alternatively, 
 in the Hill equations (which follow from Galilean transformations of the
 Newtonian equations of motion) the equivalent expression, which contains
 Oort's A constant implicitly, may be obtained by expanding the two 
 factors in --``$r \; \Omega^2$"-- about the equilibrium point and 
 keeping only linear terms in $\delta r$.  

	In the general theory of relativity, on the other hand, the effect
 of gravity is geometrically included in the equation of motion and the
 meaning of force is intrinsically different from that in the Newtonian
 theory even if the acceleration is meassured in a frame instantaneously
 at rest (Abramowicz \& Prasanna 1990, Abramowicz 1990).  Still, 
 {\it in the LRF} a procedure similar to the radial differential
 force analysis (c.f. Eq [35-37] of FT95) may be used to derive 
an approximate\footnote{ 
 Eqs [\ref{eq:Anis_Hill}] are derived under the premise that the 
 {\it standard} centrifugal and coriolis force terms act in the 
 classical sense. 
 In the Kerr geometry, a precise definition of the centrifugal force, as
 formally required to derive a relativistic analog to Eq [\ref{eq:ClasStab}],
 is hampered by the appearance of a coriolis-type radial force related 
 to the Lenz-Thirring effect (Abramowicz, Carter \& Lasota 1988).  
 These forces, which are not included in our approximate analysis,
 are important at the innermost region, $3 \ltaprx r \ltaprx 5$,
 of accretion disks in the Kerr geometry 
 (Chakrabarti \& Prasanna 1990, Prasanna \& Chakrabarti 1990).
 Their combined effect will be explored elsewhere.
} 
 relativistic analog to Eq [\ref{eq:ClasStab}].
 
 	Although geodesic Keplerian flow does not `feel' gravity, 
 deviations from relativistic Keplerian rotation unsettle a flow in 
 otherwise radial equilibrium (ignoring radial pressure gradients). 
 The equation that describes this behavior was first derived by 
 Lasota (1994, see also Abramowicz \etal 1997 and GP98)
\beq
	\half \, d_r \beta^2_r =
		- {\gamma_\phi^2 \over {\gamma_r^2} } \,
		{\sA \over {\sD}} \;\; {1\over{r^2}} 
		\left(1 - {\Omega\over{\Omega}}_+\right) 
		\left(1 - {\Omega\over{\Omega}}_-\right).
\label{eq:RadResFor} \eeq
 Recall that $\beta_r$ is the radial speed as meassured by an observer
 at rest in the corotating (CRF) frame.  

	Noting that the radial speed as meassured in the LRF by 
 definition equals zero (albeit instantaneously) we re-write the l.h.s. 
 of Eq [\ref{eq:RadResFor}] in terms of the radial gradient of this $v_r$ 
\beq
 {1 \over 2} \, d_r \beta^2_r = {1 \over 2} \, \gamma_r^{-4} \,d_r v^2_r.
\label{eq:crucial} \eeq

 	Since the gravitational potential is implicit in the geometry, 
 the r.h.s. of Eq [\ref{eq:RadResFor}] may be interpreted 
 as a Newtonian acceleration meassured in the local rest frame, 
 aka $\dot{p}_r = -\partial_r \cH ~= -\partial_r \cT$ for a 
 ``free" particle (note that the LRF is the only frame where, 
 loosely speaking, 
 it is safe to treat a `force' as a having its Newtonian meaning).

	Next, we expand the two terms in parenthesis using the definitions 
 for the prograde and retrograde Keplerian frequencies in BLF coordinates
 $\Omega_\pm = \pm (r^{3/2} \pm a)^{-1}$ 
 and Taylor expand $\Omega^2$ around either Keplerian frequency
\begin{eqnarray}
 	-\left(1 - {\Omega\over{\Omega}}_+\right) 
	 \left(1 - {\Omega\over{\Omega}}_-\right) 	\!\!\!\! &=& \!\!\!\! 
		\{1 - 2a\Omega - \Omega^2(r^3-a^2) \}	\cr
	&&\cr
 \!\!\!\!   &\simeq& \!\!\!\! 
 	\left( {1 \over {\Omega_+ \Omega_-}} \mp {a \over {\Omega_\pm}} 
	\right) \delta \Omega^2 |_{_{\rm Kep}}
\label{eq:expand} \end{eqnarray}
 where the difference $\delta\Omega^2$ is taken with respect
 to (frame-dragged) relativistic Keplerian rotation rates.
	
 	Lastly, putting together Eqs [\ref{eq:expand}, \ref{eq:crucial},
 \ref{eq:RadResFor}] to construct the radial differential force, 
 defining the relativistic shear rate in the LRF
\beq
		{\rm A}^{\rm Rel}_\Oor \equiv \sA \gamma^2    
		\left\{\half  {d\Omega \over {d \ln r}} \right\} 
\label{eq:OortsA} \eeq
 (recall this expression contains the radial gradients of $\ell$ and $\beta_r$ 
 implicitly); and noting that all other quantities in Eq [\ref{eq:ClasStab}]
 are locally defined values, the desired relativitic stability criterium 
 obtains
\beq 	-4 \left[ {1 \over {\sD}} \right] \; {1 \over {r^3}}
 	\left( {1 \over {\Omega_+ \Omega_-}} \mp {a \over {\Omega_\pm}} 
	\right) 
	{\rm A}^{\rm Rel}_\Oor \Omega_\pm 
	+ ({\rm \bf k \cdot v}_{\rm Alf})^2 < 0.
\label{eq:RelInStab} \eeq
 (recall $\Omega_- < 0$).

	The timescale derived from the relativistic modification to the 
 radial differential force follows straightforwardly by replacing
 the expression in the curly bracket of Eqs [\ref{eq:Anis_Hill}] by its
 relativistic equivalent in the l.h.s. of Ineq [\ref{eq:RelInStab}], and
 following the analysis of BH92a thereon.  
 In the limit $k_r/k_\theta \rightarrow 0$ (the horizontal regime)
 this timescale should be identified with the maximum growth rate for
 (magnetic) shearing instabilities 
\beq
	\left| { 1 \over {\tau_{_{\cM}}^{\rm Rel} }} \right|_{\rm max} =
		- \gamma^2 \left[ { \sA \over {\sD}} \right] 
		\left( \half  {d\Omega \over {d \ln r}} \right) 
		{1 \over {r^3}}
 		\left\{ r^3 - a^2  + a(r^{3 \over 2} \pm  a) \right\}. 
\label{eq:RelMRRate} \eeq
 This expression reduces to the correct limit when 
 \mbox{$a \rightarrow 0$}, with the shear force reversal at $r = 3$
 embodied by the expression for ${\rm A}^{\rm Rel}_\Oor$ as remarked
 by GP98 (see also, Anderson \& Lemos 1988, Abramowicz \& Prasanna 1990). 

 	For Keplerian, prograde orbits; using 
 Eqs [\ref{eq:ArelKep} \& \ref{eq:RelMRRate}] one finds
\beq
	-{\rm A}^{\rm eff}_{_{\cM+}} \equiv 
	\left| { 1 \over {\tau_{_\cM}^{\rm Rel} }} \right|_{\rm max} =
	{3\over4} \, \gamma^2 \left[ { \sA \over {\sD}} \right] \Omega_+,
\eeq
 while for retrograde disks, the strength of the instability is 
 somewhat diminished from its classical growth rate.

	In the horizontal regime, one may expect the maximum growth rate 
 of shear-fed instabilities (above) to differ from the relativistic 
 shear rate (compare Eqs [\ref{eq:BH_Tscal} \& \ref{eq:ArelKep}]) because
 the former corresponds (classically) to obliquous local fluid excursions
 with $\xi^x = -\xi^y$ (see \S 2.4 of BH92a) while the gradient of the
 velocity field (i.e. the shear) is maximized for excursions perpendicular 
 to $\unvc_y$.  Since the local thetrad of basis LRF vectors is rotated  
 by the geometry (c.f. \S \ref{ssec:SlimKerr}), it is not supprising that 
 new metric corrections are involved in the growth rate for 
 magnetorotational instabilities.  
 While this argument does not prove the validity of Eq [\ref{eq:RelMRRate}]
 (which is, after all, an approximate result), it does motivate our finding.





\end{document}